\documentclass[twoside,twocolumn,english,a4paper,a4paper,twocolumn,prl,twocolumn,superscriptaddress,showpacs]{revtex4-1}
\usepackage[latin9]{inputenc}
\usepackage{float}
\usepackage{amsmath}
\usepackage{graphicx}
\usepackage{amssymb}
\usepackage{float}
\usepackage{amsthm}
\usepackage{esint}
\usepackage{babel}

\begin{document}

\title{Entanglement of mechanical oscillators coupled to a non-equilibrium
environment}

\author{Max Ludwig}

\email{max.ludwig@physik.uni-erlangen.de}

\address{Department of Physics, Center for NanoScience, and Arnold Sommerfeld
Center for Theoretical Physics, Ludwig-Maximilians-Universität München,
Theresienstr. 37, D-80333 Munich, Germany}

\address{Institut für Theoretische Physik, Universität Erlangen-Nürnberg,
Staudtstrasse 7, D-91058 Erlangen, Germany}

\author{K. Hammerer}

\address{Institute for Quantum Optics and Quantum Communication,
Austrian Academy of Sciences, 
and Institute for Theoretical Physics, University of Innsbruck,Technikerstrasse 21a, 6020 Innsbruck, Austria}

\author{Florian Marquardt}

\address{Department of Physics, Center for NanoScience, and Arnold Sommerfeld
Center for Theoretical Physics, Ludwig-Maximilians-Universität München,
Theresienstr. 37, D-80333 Munich, Germany}

\address{Institut für Theoretische Physik, Universität Erlangen-Nürnberg,
Staudtstrasse 7, D-91058 Erlangen, Germany}

\address{Max Planck Institute for the Science of Light, G\"unter-Scharowsky-Str. 1/Bau 24, D-91058 Erlangen, Germany}


\begin{abstract}
Recent experiments aim at cooling nanomechanical resonators to the
ground state by coupling them to non-equilibrium environments in order
to observe quantum effects such as entanglement. This raises the general
question of how such environments affect entanglement. Here we show
that there is an optimal dissipation strength for which the entanglement
between two coupled oscillators is maximized. Our results are established
with the help of a general framework of exact quantum Langevin equations
valid for arbitrary bath spectra, in and out of equilibrium. We point
out why the commonly employed Lindblad approach fails to give even
a qualitatively correct picture. 
\end{abstract}
\maketitle

\section{1. Introduction}
Entanglement \cite{1935_Schroedinger} constitutes a cornerstone of
quantum mechanics and is a major subject of present-day research \cite{2009_Horodecki_RMPQuEntanglement}.
Whether it persists and can be observed in systems comprising macroscopic
bodies has been a hotly debated topic since the early days of quantum
mechanics. The ground state of two interacting quantum systems will
generically be entangled. Thus, one could naively expect that it is
sufficient to simply cool two interacting, macroscopic bodies to their
ground states and thereby prepare an entangled state. However, when
coupling to a dissipative bath -- as is of course necessary for cooling -- entanglement may be destroyed, as explored in a number of works,
e.g., \cite{Examples_EntanglementDissipation}. A slate of recent
experiments has now brought a new aspect into focus: A \emph{non-equilibrium}
environment, consisting of either a driven optical cavity \cite{Examples_Optomechanics},
a superconducting microwave resonator \cite{Examples_Microwave} or
a superconducting single electron transistor \cite{2006_Naik_CoolingNanomechResonator},
can be employed to cool the motion of mechanical resonators down to
the ground state. The advances in this field may ultimately enable tests of quantum mechanics in an entirely new regime  \cite{2005_Schwab} and to observe entanglement of massive objects \cite{OptomechanicsEntanglement,2008_Hartmann}.
Still it remains to resolve the issue of how the dissipative coupling
to the non-equilibrium bath affects entanglement.

In the present work, we demonstrate a non-monotonic dependence of
entanglement between two oscillators on the coupling strength to the
non-equilibrium environment and show that there is an optimal value
for the coupling to the bath. Below this value, entanglement is diminished
by thermal fluctuations, and above this value, it is lost through
dissipation. The striking behavior found here is missed entirely
by the commonly employed Lindblad approach to dissipative dynamics.

In order to obtain an exact description, we develop a general framework
based on quantum Langevin equations, which allows us to analyze the entanglement
between harmonic oscillators in the presence of coupling to a linear
bath of arbitrary spectral density. First, we exploit this scheme
to show that even in equilibrium there are effects likely to be missed
by simpler approaches. For example, the minimum coupling strength
needed for entanglement depends logarithmically on the cutoff frequency
for the most important case of an Ohmic bath spectrum. For the case
of a non-equilibrium bath, we illustrate the generic behavior in a
concrete example of two mechanical resonators inside an optical cavity,
being cooled by the optomechanical interaction with the light field
circulating in the cavity.

\begin{figure}[H]
\includegraphics[clip,width=1\columnwidth]{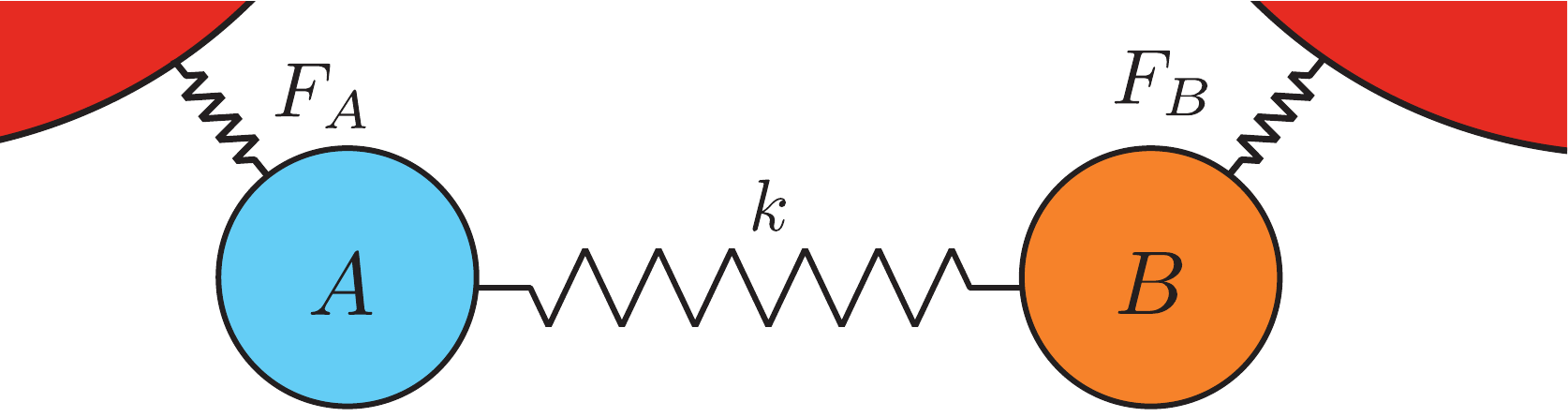}

\caption{The system consists of two harmonic oscillators ($A$ and $B$) that are
coupled via a harmonic force with spring constant $k$ and are subject
to fluctuating quantum forces ($\hat{F}_{A/B}$) due to their coupling
to the environment. }

\label{Flo:Model1} 
\end{figure}

\section{2. Model} 
We consider two coupled oscillators with masses $m_{A,B}$
and frequencies $\Omega_{A,B}$, cf. Fig.~\ref{Flo:Model1}. In terms
of their positions and momenta, $\hat{q}_{A/B}$ and $\hat{p}_{A/B},$
the Hamiltonian reads $\hat{H}_{{\rm sys}}=\sum_{\alpha=A,B}m_{\alpha}\Omega_{\alpha}^{2}\hat{q}_{\alpha}^{2}/2+\hat{p}_{\alpha}^{2}/2m_{\alpha}+k(\hat{q}_{A}-\hat{q}_{B})^{2}/2$,
with a coupling spring constant $k$. Moreover, we assume the oscillators
to be subject to fluctuating quantum forces $\hat{F}_{\alpha}$, which
are possibly correlated, and which derive from a bath of harmonic
oscillators, with $\hat{H}_{{\rm sys-bath}}=\sum_{\alpha}\hat{q}_{\alpha}\hat{F}_{\alpha}$.
They will be characterized by their spectra as specified below.

If the state of the environment is Gaussian, the oscillators also
end up in a Gaussian state, which is fully described by the covariance
matrix $\gamma_{ij}=\text{tr}(\hat{\rho}\{\hat{R}_{i},\hat{R}_{j}\}/2)$.
Here $\hat{R}=(\hat{p}_{A},\hat{q}_{A},\hat{p}_{B},\hat{q}_{B})^{T},$
$\left\langle \hat{R}_{i}\right\rangle \equiv0$ in steady state,
and $\hat{\rho}$ is the system's density matrix. As a measure of
the entanglement between the oscillators, the logarithmic negativity
\cite{1999_Eisert_EntanglementMeasures_JModOpt,2000_Simon,2002_Vidal_Entanglement}
is calculated as $E_{N}(\hat{\rho})=\sum_{i=1,2}f(\tilde{c}_{i}),$
where $f(\tilde{c})=-\log_{2}(2\tilde{c})$ for $\tilde{c}<0.5$ and
$f(\tilde{c})=0$ otherwise, and where $\tilde{c}_{1,2}$ are the
symplectic eigenvalues of the partially transposed covariance matrix
$\gamma^{T_{A}}$ \cite{2002_Vidal_Entanglement}.

\begin{figure}[h]
\includegraphics[width=1\columnwidth]{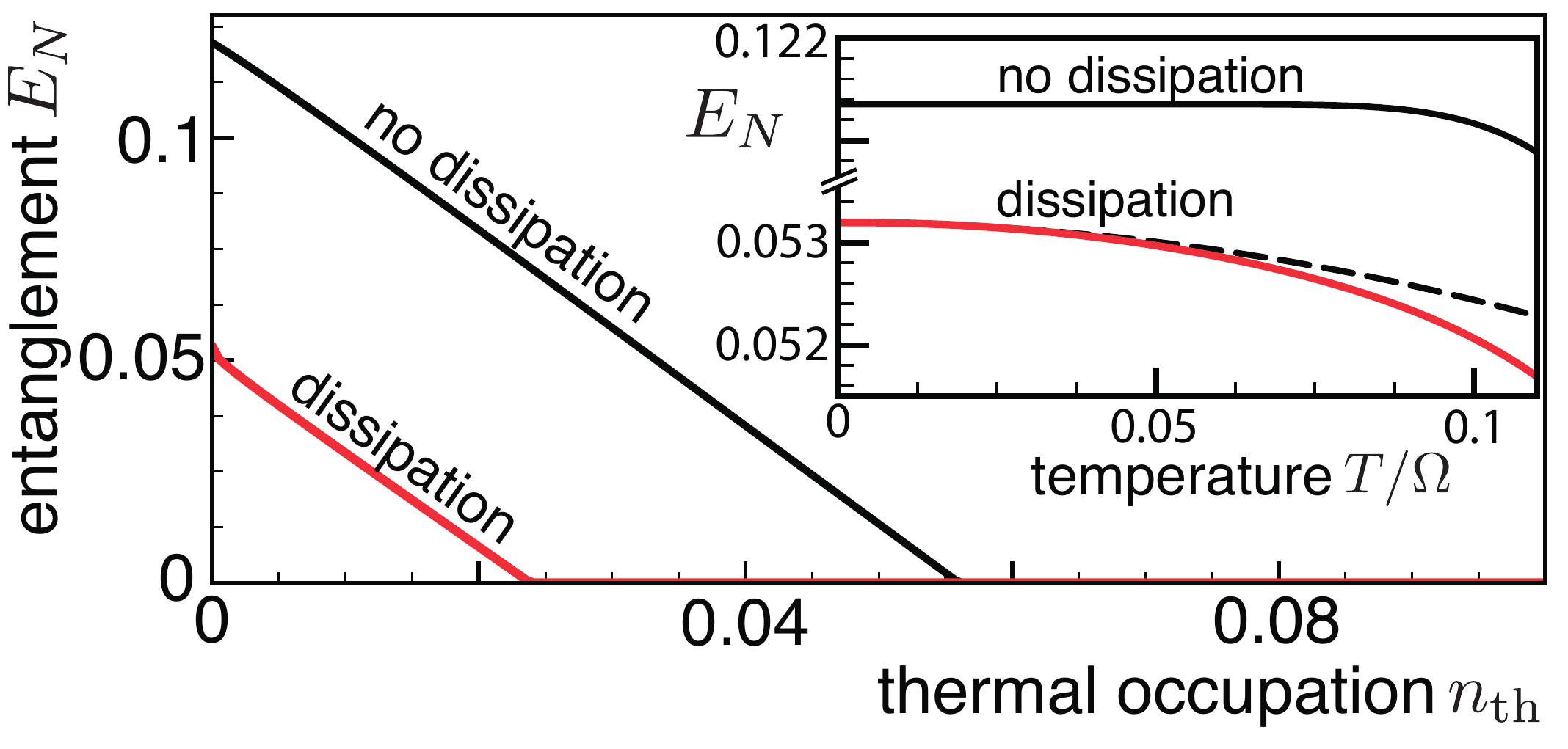}

\caption{Entanglement between two identical harmonic oscillators as a function
of the thermal occupation number $n_{{\rm th}}$ in the absence of
dissipation (black curve) and under the influence of an equilibrium
Ohmic bath (red curve), which leads to a temperature-independent reduction
of entanglement by $\Gamma_{m}\big(\ln\frac{\omega_{c}}{\Omega}-1\big)/(\pi\Omega\ln2)$.
The additional reduction with temperature scales as $T^{2}$ at low
$T$, as shown in the inset (dashed curve). $G=0.2\,\Omega,$ $\Gamma_{m}=0.1\,\Omega,$
and $\omega_{c}=10\,\Omega$.}

\label{Flo:Figure2-1} 
\end{figure}

For later use, and in order to fix the notation it will be convenient to consider first the simple example of two identical oscillators ($m_{A/B}=m,$ $\Omega_{A/B}=\Omega$) at thermal equilibrium and assume the coupling to the environment to be negligible. The system
can be decoupled by introducing the normal-mode coordinates $\hat{\eta}_{\pm}=(\hat{q}_{A}\pm\hat{q}_{B})/\sqrt{2}$
and momenta $\hat{\pi}_{\pm}=(\hat{p}_{A}\pm\hat{p}_{B})/\sqrt{2}$
corresponding to the center-of-mass motion ($\Omega_{+}=\Omega$)
and the relative motion at frequency $\Omega_{-}=\sqrt{\Omega^{2}+2k/m}\approx\Omega+2G.$
Here we defined the coupling rate $G=k/2m\Omega$, to be used in place
of $k$. In the following, for simplicity, we assume attractive interaction,
$G\propto k>0$. The symplectic eigenvalues have the simple form \begin{equation}
\tilde{c}_{1,2}=\sqrt{\left\langle \hat{\eta}_{\pm}^{2}\right\rangle \left\langle \hat{\pi}_{\mp}^{2}\right\rangle },\label{eq:simplEV}\end{equation}
 and the variances are given by $\langle\hat{\eta}_{\pm}^{2}\rangle=(2n_{\pm}+1)/2m\Omega_{\pm}$
and $\langle\hat{\pi}_{\pm}^{2}\rangle=m\Omega_{\pm}(2n_{\pm}+1)/2,$
where $n_{{\rm \pm}}=\big(e^{\Omega_{\pm}/T}-1\big)^{-1}$ is the
thermal occupation number (we set $k_{B}\equiv1$ and $\hbar\equiv1$).
Entanglement is obtained when the product $\left\langle \hat{\eta}_{-}^{2}\right\rangle \left\langle \hat{\pi}_{+}^{2}\right\rangle $
in Eq.~(\ref{eq:simplEV}) becomes smaller than $1/4$, which requires
a coupling rate $G/\Omega\gtrsim2n_{{\rm th}}$. In this expression
only terms up to first order in $G/\Omega$ have been considered and
we have set $n_{{\rm th}}=n_{+}\approx n_{-}$. For a given coupling
rate $G,$ the logarithmic negativity decreases linearly as a function
of the thermal occupation $n_{{\rm th}}$: $E_{N}(\rho)\approx(2G/\Omega-4n_{{\rm th}})/\ln2$
(black curve in Fig.~\ref{Flo:Figure2-1}). Thus, as is well known,
thermal fluctuations will reduce and eventually destroy entanglement.

\section{3. Exact solution}
Returning to the full model, an \emph{exact}
description of the dissipative dynamics is provided by quantum Langevin
equations \cite{2000_GardinerZoller_QuNoise} for the Heisenberg operators,
obtained by eliminating the bath degrees of freedom: \begin{gather}
m_{\alpha}\ddot{\hat{q}}_{\alpha}(t)+m_{\alpha}\Omega_{\alpha}^{2}\hat{q}_{\alpha}(t)+k(\hat{q}_{\alpha}(t)-\hat{q}_{\bar{\alpha}}(t))=\nonumber \\
\hat{F}_{\alpha}(t)+\intop_{-\infty}^{t}\sum_{\beta=A,B}\chi_{\alpha\beta}^{F}(t-t')\hat{q}_{\beta}(t')dt',\label{eq:LangevinEq}\end{gather}
 where $\alpha=A/B$ and $\bar{\alpha}=B/A$. $\hat{F}_{A/B}$ denotes
stationary quantum noise forces acting on the oscillators (with $\left\langle \hat{F}_{\alpha}\right\rangle =0$).
The response functions that take into account the memory effect of
the baths are given by $\chi_{\alpha\beta}^{F}(t)=-i\theta(t)\left\langle \big[\hat{F}_{\alpha}(t),\hat{F}_{\beta}(0)\big]\right\rangle $.
Solving Eq.~(\ref{eq:LangevinEq}) in Fourier space yields position
correlators $\langle\hat{q}_{\alpha}\hat{q}_{\beta}\rangle_{\omega}=\int dte^{i\omega t}\langle\hat{q}_{\alpha}(t)\hat{q}_{\beta}(0)\rangle=\sum_{\tilde{\alpha},\tilde{\beta}\in\{A,B\}}\chi_{\alpha\tilde{\alpha}}(\omega)\chi_{\beta\tilde{\beta}}(-\omega)\langle\hat{F}_{\tilde{\alpha}}\hat{F}_{\tilde{\beta}}\rangle_{\omega}$.
Here $\langle\hat{F}_{\alpha}\hat{F}_{\beta}\rangle_{\omega}=\int dte^{i\omega t}\langle\hat{F}_{\alpha}(t)\hat{F}_{\beta}(0)\rangle$
and $\chi_{\alpha\beta}(\omega)$ are elements of a matrix whose inverse
is given by $\big(\chi^{-1}(\omega)\big)_{\alpha\alpha}=m_{\alpha}(\Omega_{\alpha}^{2}-\omega^{2})+k-\chi_{\alpha\alpha}^{F}(\omega)$
and $\big(\chi^{-1}(\omega)\big)_{\alpha\beta}=-k-\chi_{\alpha\beta}^{F}(\omega)$
for $\alpha\neq\beta$. Momentum correlators follow from $\left\langle \hat{p}_{\alpha}\hat{p}_{\beta}\right\rangle _{\omega}=m_{\alpha}m_{\beta}\omega^{2}\left\langle \hat{q}_{\alpha}\hat{q}_{\beta}\right\rangle _{\omega}$,
and $\left\langle \hat{p}_{\alpha}\hat{q}_{\beta}\right\rangle _{\omega}=-im_{\alpha}\omega_{\alpha}\left\langle \hat{q}_{\alpha}\hat{q}_{\beta}\right\rangle _{\omega}$.
Finally, equal-time correlators are obtained by integration, $\left\langle \hat{q}_{\alpha}\hat{q}_{\beta}\right\rangle =\int\frac{d\omega}{2\pi}\left\langle \hat{q}_{\alpha}\hat{q}_{\beta}\right\rangle _{\omega}$.
The solution of Eq.~(\ref{eq:LangevinEq}) thus provides the full
covariance matrix $\gamma$ in terms of frequency integrals over arbitrary
bath spectra, and from it the logarithmic negativity $E_{N}$ for
two coupled dissipative oscillators. Note that we did not assume equilibrium,
i.e., the fluctuation-dissipation relation between $\chi_{\alpha\beta}^{F}$
and $\left\langle \hat{F}_{\alpha}\hat{F}_{\beta}\right\rangle _{\omega}$
does not necessarily hold.

For simplicity, we will from now on restrict our explicit calculations
to the symmetric case of two identical oscillators that couple equally
strongly to independent baths, such that $\langle\hat{F}_{\alpha}\hat{F}_{\beta}\rangle_{\omega}=\delta_{\alpha\beta}\langle\hat{F}\hat{F}\rangle_{\omega}$
and $\chi_{\alpha\beta}^{F}=\delta_{\alpha\beta}\chi^{F}.$ The system
can then, as before, be decomposed into the center of mass mode ($\hat{\eta}_{+}$,
$\hat{\pi}_{+}$) and the relative mode ($\hat{\eta}_{-}$, $\hat{\pi}_{-}$),
which become independent dissipative oscillators. We find $\left\langle \hat{\eta}_{\pm}\hat{\eta}_{\pm}\right\rangle _{\omega}=\left\langle \hat{F}\hat{F}\right\rangle _{\omega}\left|\chi_{\pm}(\omega)\right|^{2}$,
where $\chi_{\pm}(\omega)=\big(m(\Omega_{\pm}^{2}-\omega^{2})-\chi^{F}(\omega)\big)^{-1}$.
After frequency integration, Eq.~(\ref{eq:simplEV}) thus directly
yields the logarithmic negativity. %

\section{4. Equilibrium bath}
First, we illustrate the general scheme
for the case of equilibrium baths, picking the important example of
an Ohmic bath spectrum: $\langle\hat{F}_{\alpha}\hat{F}_{\alpha}\rangle_{\omega}^{T}=\langle\hat{F}\hat{F}\rangle_{\omega}^{T}=m\Gamma_{m}\omega(\coth(\omega/2T)+1)/(1+\omega^{2}/\omega_{c}^{2})$.
Here $T$ denotes the temperature, $\Gamma_{m}$ the damping rate,
and $\omega_{c}$ the cutoff frequency. For $\Gamma_{m}<\Omega$ and
$\omega_{c}\gg\Omega$, the position and momentum variances of an
oscillator coupled to this bath are given analytically in \cite{1992_Weiss_QuDissipativeSystems}.
Here we only display the expansion to first order in $\Gamma_{m}/\Omega$
at $T=0$:\begin{eqnarray}
2m\Omega_{\pm}\left\langle \hat{\eta}_{\pm}^{2}\right\rangle  & \approx & 1-\frac{\Gamma_{m}}{\pi\Omega_{\pm}}\nonumber \\
2\left\langle \hat{\pi}_{\pm}^{2}\right\rangle /m\Omega_{\pm} & \approx & 1+\frac{\Gamma_{m}}{\pi\Omega_{\pm}}\big(2\ln\frac{\omega_{c}}{\Omega_{\pm}}-1\big).\label{eq:correlator4}\end{eqnarray}

As illustrated in Fig.~\ref{Flo:Figure2-1}, entanglement between
the oscillators is suppressed due to their coupling to the bath. The
high-frequency bath modes cause momentum fluctuations that depend
logarithmically on the cutoff frequency, cf. Eq.~(\ref{eq:correlator4}).
Thus, even at zero temperature, the coupling to the environment reduces
the logarithmic negativity by $\Gamma_{m}\big(\ln\frac{\omega_{c}}{\Omega}-1\big)/(\pi\Omega\ln2)$
{[}as follows from Eqs.~(\ref{eq:simplEV}) and (\ref{eq:correlator4}){]},
and eventually destroys the entanglement completely. Entanglement
persists ($E_{N}>0$) only if the coupling rate exceeds a threshold
value of

\begin{equation}
G_{{\rm min}}^{{\rm Ohmic},T=0}=\frac{\Gamma_{m}}{\pi}\big(\ln\frac{\omega_{c}}{\Omega}-1\big)\,.\label{eq:G_ohmic_min}\end{equation}
 As a distinctive feature, the minimal coupling rate depends logarithmically
on the cutoff frequency. It indicates that any approach that disregards
the influence of high-frequency fluctuations has to fail, as discussed
for the example of the Lindblad approach below. Our general formula
also allows to obtain the full temperature-dependence, cf. Fig. \ref{Flo:Figure2-1}.

\section{5. Non-equilibrium bath} 
\begin{figure}[h]
\includegraphics[width=1\columnwidth]{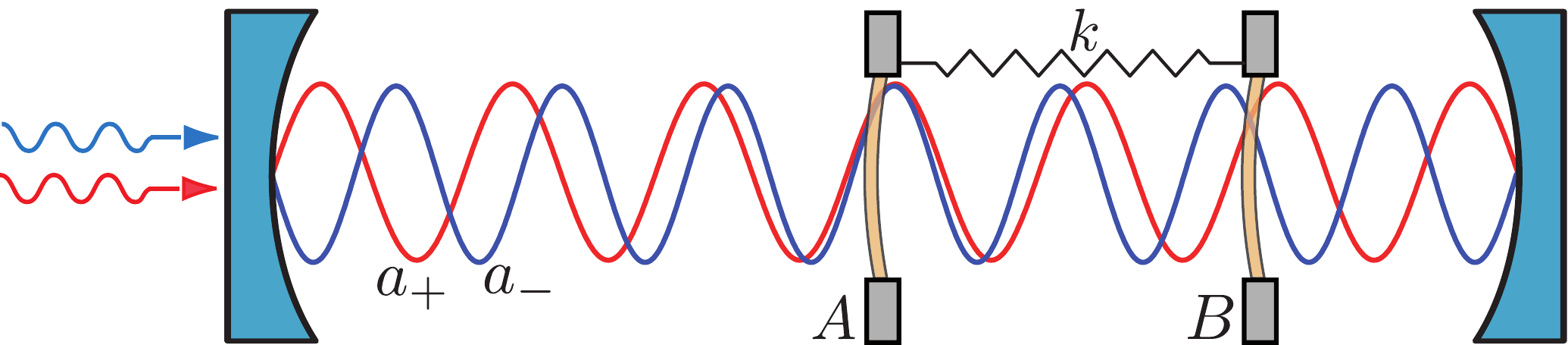}

\caption{Two coupled mechanical oscillators, represented by membranes, under
the influence of non-equilibrium photon shot noise baths because of their
coupling to two modes ($a_{\pm}$) of an optical cavity. The placement
of the membranes allows the normal modes of the coupled system to be cooled
by independent noise forces $F_{\pm}^{{\rm cav}}$.}

\label{Flo:Model2} 
\end{figure}

Tunable non-equilibrium quantum fluctuations
are now relevant in many contexts, and may be used, e.g., to cool
systems below the bulk temperature. A paradigmatic example is the
photon shot noise coupled to mechanical resonators in optomechanical
setups \cite{Examples_Optomechanics,2009_FM_ReviewOptomechanics}
(the following results also apply to analogous electromechanical systems
\cite{Examples_Microwave,2006_Naik_CoolingNanomechResonator}). We
treat the conceptually clearest case where two nanomechanical membranes
are placed inside a laser-driven cavity and two independent light
forces $\hat{F}_{\pm}^{{\rm cav}}$ act on the mechanical normal modes
$\hat{\eta}_{\pm}$ leading to optomechanical cooling \cite{2007_FM_SidebandCooling,2007_Wilson-Rae_TheoryGroundStateCooling}.
This may be realized in a setup with two cavity modes, where $\hat{H}_{{\rm sys-cav}}=(g/\ell_{m})\big((\hat{a}_{+}^{\dagger}+\hat{a}_{+})\hat{\eta}_{+}+(\hat{a}_{-}^{\dagger}+\hat{a}_{-})\hat{\eta}_{-}\big)$,
cf. Fig. \ref{Flo:Model2}. Here $\hat{a}_{\pm}$ are the annihilation
operators of the cavity modes, $\ell_{m}=1/\sqrt{2m\Omega}$ is the
mechanical ground-state width, and $g$ is the oscillator-cavity coupling
rate that scales linearly with the laser amplitude (see \cite{2009_Hammerer_StrongCoupling,2010_Wallquist_PRA}
for a derivation of this type of coupling). The mechanical coupling
$k$ between the oscillators (here assumed as given) can itself be
implemented via other, strongly driven far-detuned cavity modes \cite{2008_Hartmann,2009_Hammerer_StrongCoupling}.
Other possible setups include cold-atom or hybrid atom-membrane systems
\cite{2009_Hammerer_StrongCoupling}.

Elimination of the cavity degrees of freedom generates cavity noise spectra
\cite{2007_FM_SidebandCooling} $\left\langle \hat{F}_{\pm}\hat{F}_{\pm}\right\rangle _{\omega}^{{\rm cav}}=(g/\ell_{m})^{2}\kappa[(\omega+\Delta_{\pm})^{2}+\kappa^{2}/4]^{-1}$,
where $\kappa$ is the decay rate of the cavity photons and $\Delta_{\pm}$
the detuning of the corresponding input lasers with respect to the
first (second) cavity mode. A spectrum of this kind induces an optomechanical
cooling rate of $\Gamma_{{\rm opt,\pm}}=\ell_{m}^{2}\Big(\left\langle \hat{F}_{\pm}\hat{F}_{\pm}\right\rangle _{\Omega_{\pm}}^{{\rm cav}}-\left\langle \hat{F}_{\pm}\hat{F}_{\pm}\right\rangle _{-\Omega_{\pm}}^{{\rm cav}}\Big)$.
In the optimal cooling regime, for $\Delta_{\pm}=-\Omega_{\pm}$ and
$\kappa\ll\Omega$, we have $\Gamma_{{\rm opt}\pm}\approx\Gamma_{{\rm opt}}=4g^{2}/\kappa$.
In this regime, the minimum possible phonon number due to optical
cooling, defined by $(n_{{\rm opt}\pm}+1)/n_{{\rm opt}\pm}=\left\langle \hat{F}_{\pm}\hat{F}_{\pm}\right\rangle _{\Omega_{\pm}}^{{\rm cav}}/\left\langle \hat{F}_{\pm}\hat{F}_{\pm}\right\rangle _{-\Omega_{\pm}}^{{\rm cav}}$,
will be much smaller than~1 ($n_{{\rm opt}\pm}\approx n_{{\rm opt}}=(\kappa/4\Omega)^{2}$).
Moreover, we assume $\Gamma_{m}n_{{\rm th}}\ll\Omega,$ $\Gamma_{m}\ll\Gamma_{{\rm opt}}$
and $g\ll\Omega$ as required for ground state cooling. The full forces
$\hat{F}_{\pm}=\hat{F}_{\pm}^{{\rm cav}}+\hat{F}_{\pm}^{T}$ also
contain thermal fluctuations, $\hat{F}_{\pm}^{T}$, independent from
$\hat{F}_{\pm}^{{\rm cav}}$. For low mechanical damping ($\Gamma_{m}\ll\Omega$)
the spectrum of the thermal bath can be replaced by the values at
the resonances, i.e., $\langle\hat{F}_{\pm}\hat{F}_{\pm}\rangle_{\omega}^{T}\mapsto\langle\hat{F}_{\pm}\hat{F}_{\pm}\rangle_{\omega={\rm sgn(\omega)}\Omega_{\pm}}^{T}$.
The general scheme yields the variances by integrating $\langle\hat{\eta}_{\pm}\hat{\eta}_{\pm}\rangle_{\omega}=(\langle\hat{F}_{\pm}\hat{F}_{\pm}\rangle_{\omega}^{T}+\langle\hat{F}_{\pm}\hat{F}_{\pm}\rangle_{\omega}^{{\rm cav}})\mid\chi_{\pm}(\omega)\mid^{2}$
and $\langle\hat{\pi}_{\pm}\hat{\pi}_{\pm}\rangle_{\omega}=m^{2}\omega^{2}\langle\hat{\eta}_{\pm}\hat{\eta}_{\pm}\rangle_{\omega}$.

In the optimal cooling regime, the variances of the optomechanically
damped system can be expressed in a compact way:\begin{eqnarray}
2\langle\hat{\pi}_{\pm}^{2}\rangle/m\Omega_{\pm} & \approx & 1+2(n_{{\rm eff}}+\delta n),\label{eq:correlator5p}\\
2m\Omega_{\pm}\langle\hat{\eta}_{\pm}^{2}\rangle & \approx & 2\langle\pi_{\pm}^{2}\rangle/m\Omega_{\pm}+g^{2}/\Omega_{\pm}^{2},\label{eq:correlator5x}\end{eqnarray}
 where $n_{{\rm eff}}=\Gamma_{m}n_{{\rm th}}/\Gamma_{{\rm opt}}+n_{{\rm opt}}$
and $\delta n=\Gamma_{m}n_{{\rm th}}/\kappa$.

Together with Eq.~(\ref{eq:simplEV}), these formulas constitute
our main result for entanglement in a system subject to optomechanical
cooling. We now extract and discuss its main physical features. The
first term on the rhs. of Eq.~(\ref{eq:correlator5p}) describes
the ground state energy, and the second term takes account of the cooling
mechanism: the thermal occupation is reduced to an effective phonon
number $n_{{\rm eff}}$. Thus, entanglement can in principle be created
even for large bulk temperatures, $n_{{\rm th}}\gg1$, if the optomechanical
damping rate $\Gamma_{{\rm opt}}$ is sufficiently large. Since $\Gamma_{{\rm opt}}=4g^{2}/\kappa$,
this can be achieved either by reducing the cavity linewidth $\kappa$
or by increasing the cavity-oscillator coupling rate $g$. However,
we identify two processes that destroy entanglement for small $\kappa$
and large $g$, respectively. First, as known from \cite{2007_FM_SidebandCooling},
the cooling mechanism becomes less efficient in the strong coupling
regime $\Gamma_{{\rm opt}}\sim\kappa$, where the contribution of
$\delta n$ becomes appreciable. Second, for a large optomechanical
coupling strength $g$, the low-frequency contributions of the photon
shot noise induce an increase of the position variance {[}second term
on the rhs. of Eq.~(\ref{eq:correlator5x}){]}. This implies that
strong correlations between the individual oscillators and the driven
cavity lead to a destruction of entanglement between the oscillators.%
\begin{figure}[h]
\includegraphics[width=0.8\columnwidth]{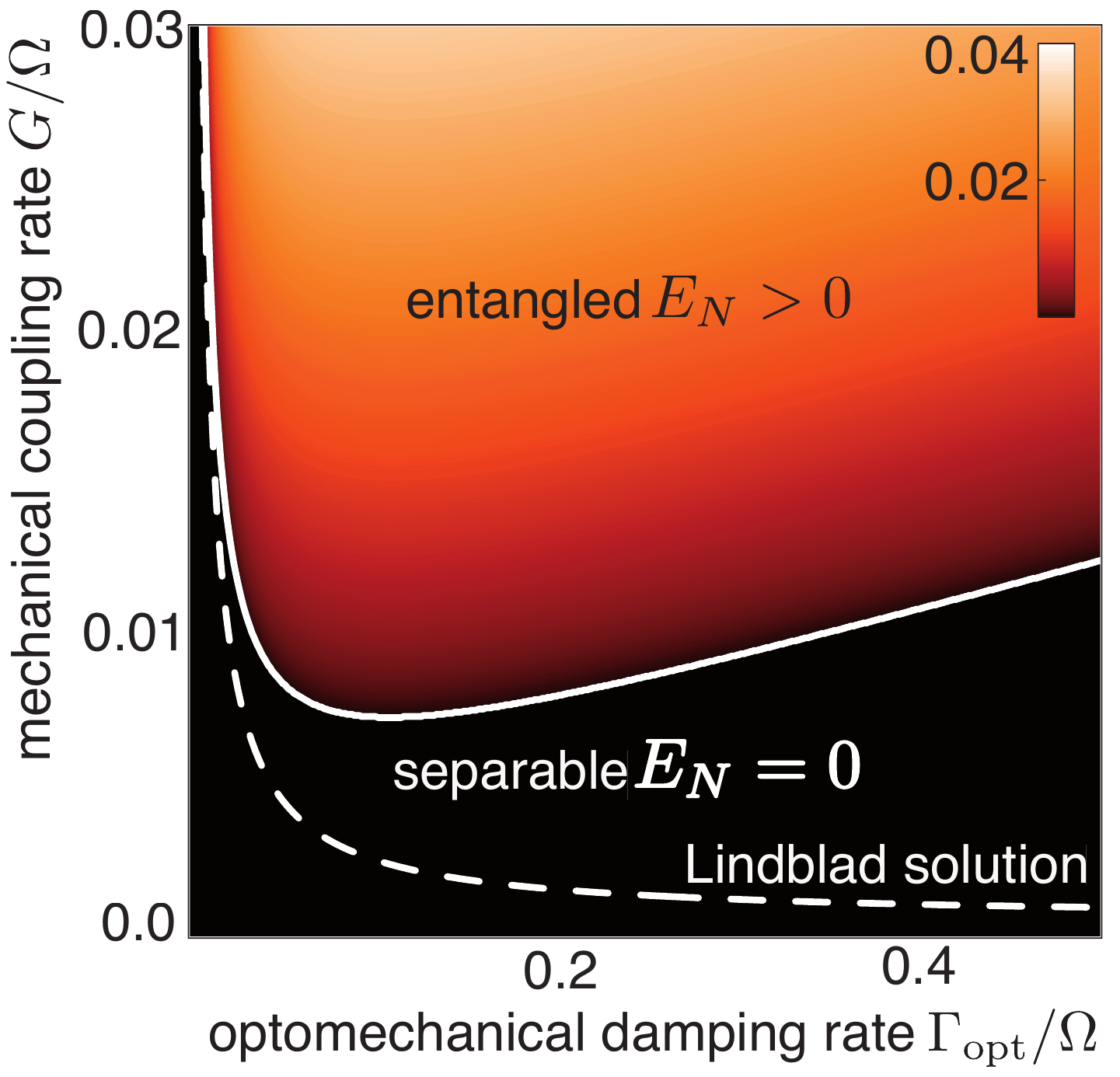}

\caption{Density plot of the entanglement (logarithmic negativity $E_{N}$)
for the non-equilibrium dissipative system of Fig.~\ref{Flo:Model2},
as a function of the optomechanical cooling rate $\Gamma_{{\rm opt}}$
and the mechanical coupling rate $G$. The white solid line represents
the boundary between entangled and separable states of the system
and thereby defines the minimal coupling rate $G_{{\rm min}}$ necessary
to observe entanglement. The dashed white line depicts the result
for $G_{{\rm min}}$ from the simpler Lindblad approach. $\Gamma_{m}n_{{\rm th}}=10^{-4}\,\Omega,$
$\Delta_{\pm}=-\Omega_{\pm}$ and $\kappa=0.067\,\Omega$.}

\label{Flo:Figure2} 
\end{figure}

\begin{figure}[h]
\includegraphics[width=0.8\columnwidth]{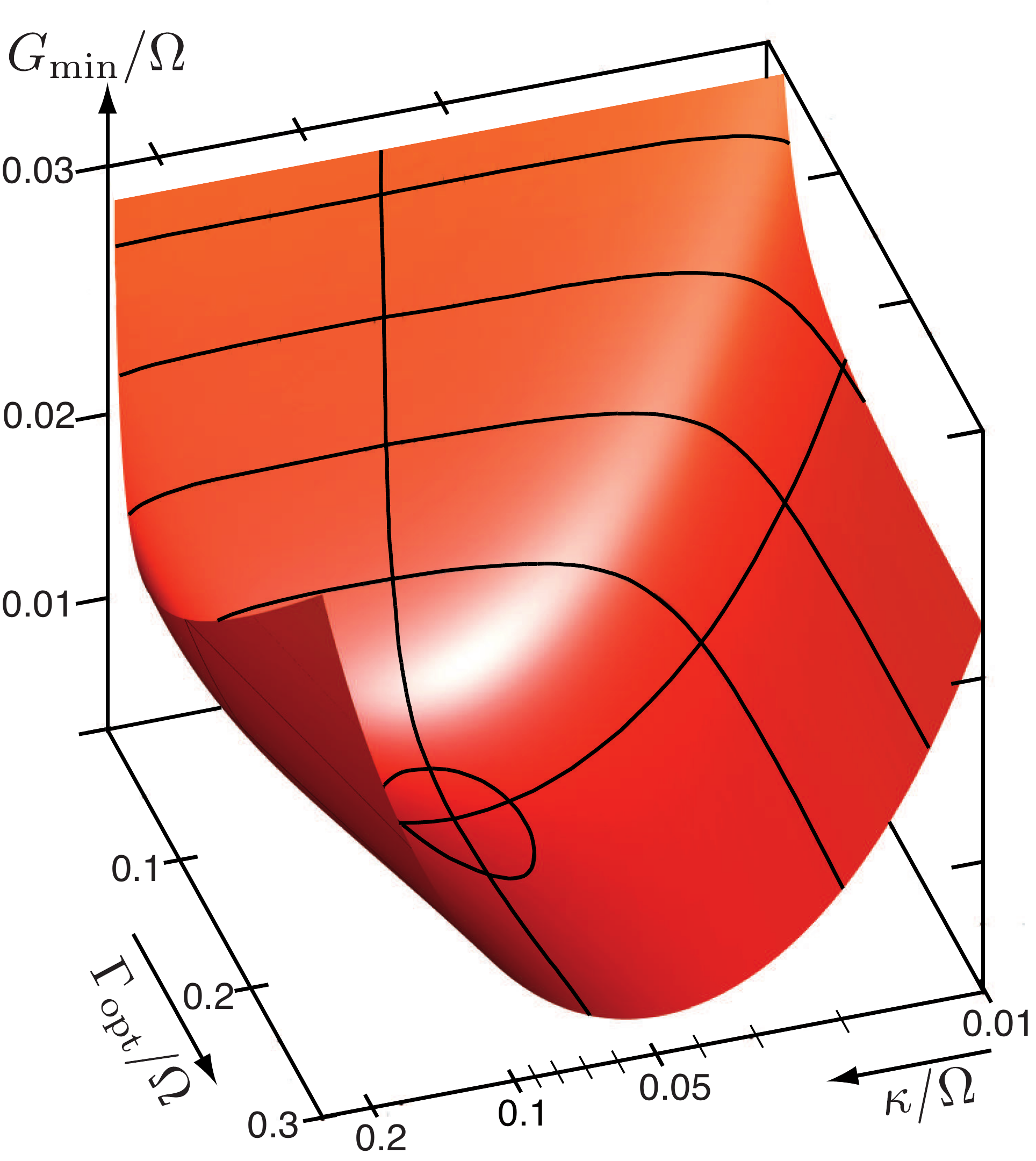}

\caption{Minimal coupling strength $G_{{\rm min}}$ necessary to observe entanglement.
$G_{{\rm min}}$ displays an optimum for intermediate values of
the optomechanical damping rate $\Gamma_{{\rm opt}}$ and the cavity
decay rate $\kappa$ (for $\Gamma_{m}n_{{\rm th}}=10^{-4}\,\Omega,$
$\Delta_{\pm}=-\Omega_{\pm}$). }

\label{Flo:Figure3} 
\end{figure}

As a consequence, entanglement depends non-monotonically on the cavity
linewidth $\kappa$ and the optomechanical damping rate $\Gamma_{{\rm opt}}$
in the optimal cooling regime, cf. Figs.~\ref{Flo:Figure2} and \ref{Flo:Figure3}.
Entanglement can be generated only if the mechanical coupling rate
exceeds a value of \begin{equation}
G_{{\rm min}}/\Omega\approx2(n_{{\rm eff}}+\delta n)+\Gamma_{{\rm opt}}\kappa/8\Omega^{2}.\label{eq:G_opt_cool}\end{equation} 
Note that Eq.~(\ref{eq:G_opt_cool}) can be employed to optimize
entanglement.

\section{6. Shortcomings of the Lindblad approach}
The crucial destruction of entanglement by strong dissipation is missed entirely by the commonly
employed Lindblad master equation approach. Its general form is given
by $\dot{\hat{\rho}}=-i[\hat{H}_{{\rm sys}},\hat{\rho}]+\sum_{i}\mathcal{L}_{i}(\hat{\rho})$,
where the influence of the bath is taken into account by Lindblad
terms $\mathcal{L}_{i}$ \cite{2000_GardinerZoller_QuNoise}. For
equilibrium baths, these are given by $\mathcal{L}_{m,\downarrow}^{(\pm)}(\hat{\rho})=(\Gamma_{m}/2)(n_{\pm}+1)\mathcal{D}[\hat{A}_{\pm}]$
and $\mathcal{L}_{m,\uparrow}^{(\pm)}(\hat{\rho})=(\Gamma_{m}/2)n_{\pm}\mathcal{D}[\hat{A}_{\pm}^{\dagger}],$
where $\mathcal{D}[\hat{A}](\hat{\rho})=2\hat{A}\hat{\rho}\hat{A}^{\dagger}-\hat{A}^{\dagger}\hat{A}\hat{\rho}-\hat{\rho}\hat{A}^{\dagger}\hat{A}$
and $\hat{A}_{\pm}$ are the mechanical normal mode annihilation operators.
At zero temperature, the shortcomings of the Lindblad approach are
most obvious: The system evolves into its ground state, whose entanglement
is not reduced at all by the system-bath coupling. To treat the non-equilibrium
case of Fig.~\ref{Flo:Model2} in the Lindblad approach, we have
to consider four additional terms, $\mathcal{L}_{c,\downarrow}^{(\pm)}(\hat{\rho})=\left\langle F_{\pm}F_{\pm}\right\rangle _{\Omega_{\pm}}^{{\rm cav}}(\Omega_{\pm}\ell_{m}^{2}/2\Omega)\mathcal{D}[\hat{A}_{\pm}](\hat{\rho})$
and $\mathcal{L}_{c,\uparrow}^{(\pm)}(\hat{\rho})=\left\langle F_{\pm}F_{\pm}\right\rangle _{-\Omega_{\pm}}^{{\rm cav}}(\Omega_{\pm}\ell_{m}^{2}/2\Omega)\mathcal{D}[\hat{A}_{\pm}^{\dagger}](\hat{\rho})$,
which take account of the decoherence via the cavity modes (see \cite{2010_Wallquist_PRA}
for a detailed derivation). The steady-state variances of the normal
modes follow as $ $$2m\Omega_{\pm}\langle\hat{\eta}_{\pm}^{2}\rangle=2\langle\hat{\pi}_{\pm}^{2}\rangle/m\Omega_{\pm}\approx2n_{{\rm eff}}+1.$
This expression describes the cooling to an effective phonon number
$n_{{\rm eff}}$ but fails to capture the loss of entanglement for
strong optomechanical coupling (see the dashed curve in Fig.~\ref{Flo:Figure2}).
The shortcomings of this approach can be understood by noting that
the Born-Markov approximation, which assumes the bath to have a very
short correlation time (no memory) and to be uncorrelated with respect
to the system, does not hold in general for a non-equilibrium bath
as can be seen in our example.

\section{7. Conclusions and outlook}
The general exact framework introduced
here can be employed to analyze the entanglement of oscillators under
the influence of arbitrary bath spectra, among them non-equilibrium
and tailored non-standard spectral densities. As pointed out in this
paper, the effects of tunable non-equilibrium environments promise
rich physics to be explored in current experimental setups. The optomechanical
setup investigated here is in fact just one of a rather large class
of setups to which this work applies, and which also extends into
the fields of electromechanics \cite{Examples_Microwave,2006_Naik_CoolingNanomechResonator}
and cold-atom physics \cite{2009_Hammerer_StrongCoupling}. We also
note that completely different systems show similar entanglement production
effects under non-equilibrium conditions, as has been explored in
the case of coupled, driven qubits \cite{2009_Li_EntanglementInDrivenQubitSystems},
atoms \cite{EntanglementProduction_Atoms} and ions \cite{2002_Schneider_EntanglementInDickeModel}, or coupled double quantum
dots \cite{2007_Lambert_NoneqEntanglement}.

In the quest to observe entanglement in dissipatively cooled optomechanical or nanoelectromechanical systems the theory presented here serves as an essential guideline: It identifies viable parameter regimes for generating and optimizing entanglement between massive mechanical oscillators.

Recent works \cite{Entanglement_ParametricAmplification,2010_Wallquist_PRA} have proposed an alternative way of generating entanglement in nanomechanical systems: By modulation the coupling strength between the oscillators the system can be parametrically driven into a non-equilibrium state which features entanglement even at relatively large temperatures. In a future work the general framework introduced here can be employed to discuss the generation of entanglement in a parametrically driven system and to compare and connect the two approaches.

\section{Acknowledgments} 
Support by the DFG through NIM, SFB631 and
the Emmy-Noether program, by the Austrian Science Fund through SFB
FOQUS, and by the IQOQI is acknowledged.

\bibliographystyle{plain}
\bibliography{bib}

\end{document}